\documentclass[a4paper,11pt]{article}
\usepackage{pos}
\renewcommand{\logo}{\relax}

\title{Probing  \boldmath{$\tau$} lepton dipole moments at future Muon Colliders}

\author*[a]{ZeQiang Wang}
\affiliation[a]{Centre for Cosmology, Particle Physics and Phenomenology (CP3),
Université catholique de Louvain, 1348 Louvain-la-Neuve, Belgium}
\emailAdd{zeqiang.wang@uclouvain.be}

\abstract{The anomalous magnetic moments of leptons represent excellent probes of the Standard Model and therefore also of possible new physics effects.
In particular, the persisting hint of new physics in the muon $g$-2 motivates the investigation of similar effects also in the other leptonic dipoles.
In this work, we examine the new physics sensitivity of the tau $g$-2 at future high-energy Muon Collider. We show that these facilities can access a number of processes like the Drell-Yan processes 
$\mu^+\mu^- \to \tau^+\tau^-(h)$, or vector-boson-fusion processes such 
as $\mu^+\mu^- \to \mu^+\mu^-\tau^+\tau^-$ and $\mu^+\mu^-\to \nu_\mu \bar{\nu}_\mu  \tau^+\tau^-$ those can probe the tau $g$-2
at the level of $\mathcal{O}(10^{-5}-10^{-4})$, a resolution that is orders 
of magnitude better than the current bounds.}

\FullConference{42nd International Conference on High Energy Physics (ICHEP2024)\\
18-24 July 2024\\
Prague, Czech Republic\\}

\begin{document}
\renewcommand{\hookAfterAbstract}{
\par\bigskip
\textsc{IRMP-CP3-24-30}
}
\maketitle
\section{Introduction} 
The anomalous magnetic moment of the muon has provided, over the last ten years, an enduring hint for 
new physics (NP). The comparison between the Standard Model prediction  of $a_\mu = (g_\mu-2)/2$~\cite{Aoyama:2020ynm} and the experimental 
value of the E821 experiment at BNL~\cite{Bennett:2006fi} reveals a interesting $\sim 3.7 \,\sigma$ discrepancy. The Fermilab  Run-1/2/3 presented here, $\Delta a_{\mu}$(FNAL) show a  $5.0 \,\sigma$ deviation from the 
2020 SM prediction~\cite{Aoyama:2020ynm,Muong-2:2023cdq}.
The latest experimental world average 
combined (BNL and FNAL) experimental (exp) average is $a_{\mu}(\text{exp}) = 116\,592\,059(22) \times 10^{-11}$ (0.19 ppm), which represents a factor of 2 improvement in precision~\cite{Muong-2:2023cdq}. Based on these results, a precise measurement of $a_{\tau}$ would offer a new excellent opportunity to unveil NP effects. Recently, the CMS collaboration has observed, for the first time, $\gamma \gamma \to \tau^+\tau^- $ events in pp runs, which improved previous constrains on tau $g-2$ by a factor of 5 ~\cite{CMS:2024qjo}. 
Indeed, many NP theories predict dominant couplings to third families and in such a case the present discrepancy 
in $\Delta a_{\mu}$ would suggest NP effects in $\Delta a_{\tau}$ much larger than the value of $\mathcal{O}(10^{-6})$ 
predicted by naive scaling $\Delta a_{\tau} / \Delta a_{\mu} = m^2_\tau/m^2_{\mu}$~\cite{Giudice:2012ms}. 
\section{Leptonic $g$-2 in the Standard Model EFT} 
New interactions emerging at a scale $\Lambda$ larger than the electroweak scale can be described at energies $E \ll \Lambda$ 
by an effective Lagrangian containing non-renormalizable $SU(3)_c \otimes SU(2)_L \otimes U(1)_Y$ invariant operators.
Focusing on the leptonic $g$-2, the relevant effective Lagrangian contributing to them, up to one-loop order, reads
%
\begin{align}
\mathcal{L} &= 
\frac{C^\ell_{eB}}{\Lambda^2}
\left( \bar\ell_L \sigma^{\mu\nu}e_{R}\right) \! H B_{\mu\nu} + 
\frac{C^\ell_{eW}}{\Lambda^2}
\left( \bar\ell_L \sigma^{\mu\nu} e_{R} \right) \! \tau^I \! H W_{\mu\nu}^I 
\nonumber\\
&+ \frac{C^\ell_{T}}{\Lambda^2}( \overline{\ell}^a_L\sigma_{\mu\nu}e_{R}) \varepsilon_{ab} (\overline{Q}^b_L\sigma^{\mu\nu} u_{R}) 
+ h.c.
\label{eq:L_SMEFT}
\end{align} 
%
where it is assumed that the NP scale $\Lambda \gtrsim 1$ TeV. 

Starting from Eq.~(\ref{eq:L_SMEFT}), the resulting expression for $\Delta a_\tau$ is given by
%
\begin{align}
\Delta a_\tau  &\simeq \frac{4m_\tau v}{e\sqrt{2}\Lambda^2} \, 
\bigg(
C^\tau_{e\gamma} - \frac{3\alpha}{2\pi} \frac{c^2_W \!-\! s^2_W}{s_W c_W} \,C^\ell_{eZ} \log\frac{\Lambda}{m_Z}
\bigg)
\nonumber\\
& - \frac{4m_\tau m_t}{\pi^2} \frac{C_T^{\tau t}}{\Lambda^2}\,
\log\frac{\Lambda}{m_t},
\label{eq:Delta_a_ell}
\end{align}
%
where $s_W$($c_W$) is the sine (cosine) of the weak mixing angle, $C_{e\gamma}=c_W C_{eB} - s_W C_{eW}$ 
and $C_{eZ} = -s_W C_{eB} - c_W C_{eW}$.
In order to see where we stand, let us determine the NP scale probed by $\Delta a_\tau$.
From eq.~(\ref{eq:Delta_a_ell}) we find that
\begin{align}
\!\!\!\Delta a_\tau & \! \approx 
4 \!\times\! 10^{-5} \! \left(\frac{10 \,{\rm TeV}}{\Lambda}\right)^{\!\!2} \!\!\!
\left(C^\tau_{e\gamma} \!-\! 0.12 \, C^{\tau t}_T \!-\! 0.02 \, C^{\tau}_{eZ}\right) 
\end{align}
where $C^\tau_{e\gamma}$ in the above expression has been evaluated at the scale $m_\tau$ by including its one-loop renormalization effects and we neglected subleading contributions from the charm quark stemming from $C^{\tau c}_T$.
\section{Results}
We now compute the reach on the dipole operators from the Drell-Yan processes 
$\mu^+\mu^- \to \tau^+\tau^-(h)$, or vector-boson-fusion processes such 
as $\mu^+\mu^- \to \mu^+\mu^-\tau^+\tau^-$ and $\mu^+\mu^-\to \nu_\mu \bar{\nu}_\mu  \tau^+\tau^-$ process at a high-energy muon collider, simulating the signal and background processes in MadGraph at parton level. We consider  an 80\% efficiency for tau identification, and a 50\% efficiency for the reconstruction of a boosted Higgs decaying into $b\bar b$. 
We impose basic acceptance cuts $\eta < 2.5$ for all final state particles (including the boosted Higgs boson), and further require $\Delta R_{\tau\tau} > 0.4$. In order to suppress the SM backgrounds we impose the following analysis cuts:
\begin{equation}
p_{T,\tau} > E_{\rm beam}/10,\qquad p_{T,h} > E_{\rm beam}/10,\qquad M_{\tau\tau} > E_{\rm beam}/10,
\end{equation}

\begin{figure}
    \centering%
    \begin{minipage}[b]{0.22\textwidth}
        \centering%
        \includegraphics[width=\textwidth]{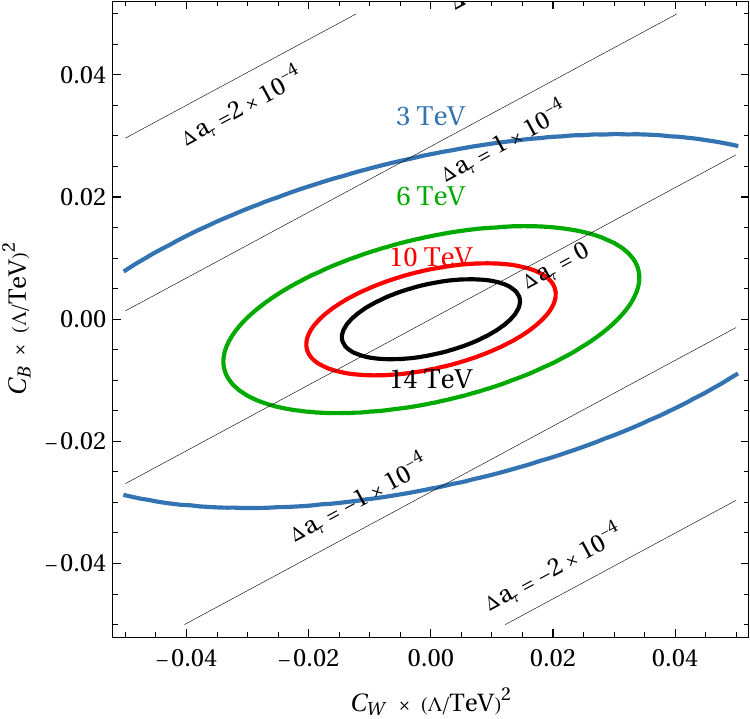}
        \caption*{(a) $\mu^+\mu^-\to \tau^+\tau^- $} 
    \end{minipage}%
    \hfill%
    \begin{minipage}[b]{0.22\textwidth}
        \centering%
        \includegraphics[width=\textwidth]{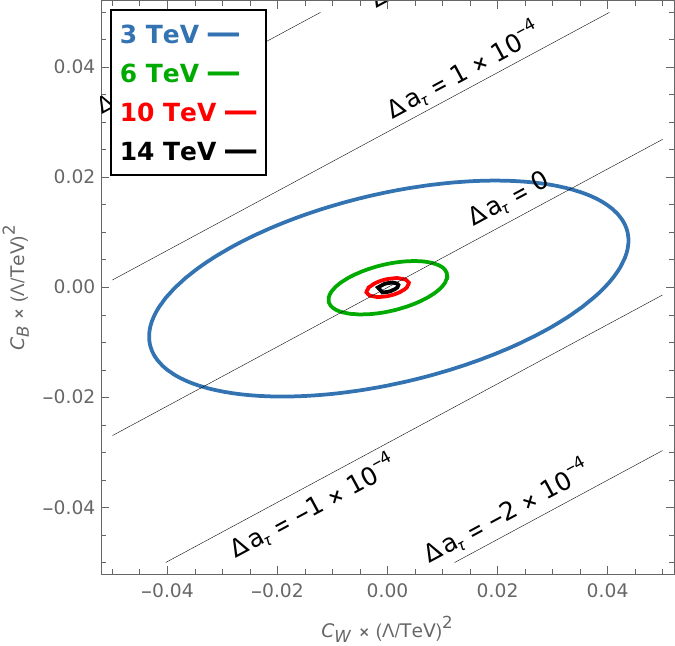}
        \caption*{(b) $\mu^+\mu^-\to \tau^+\tau^- h$}
    \end{minipage}%
    \hfill%
    \begin{minipage}[b]{0.22\textwidth}
        \centering%
        \includegraphics[width=\textwidth]{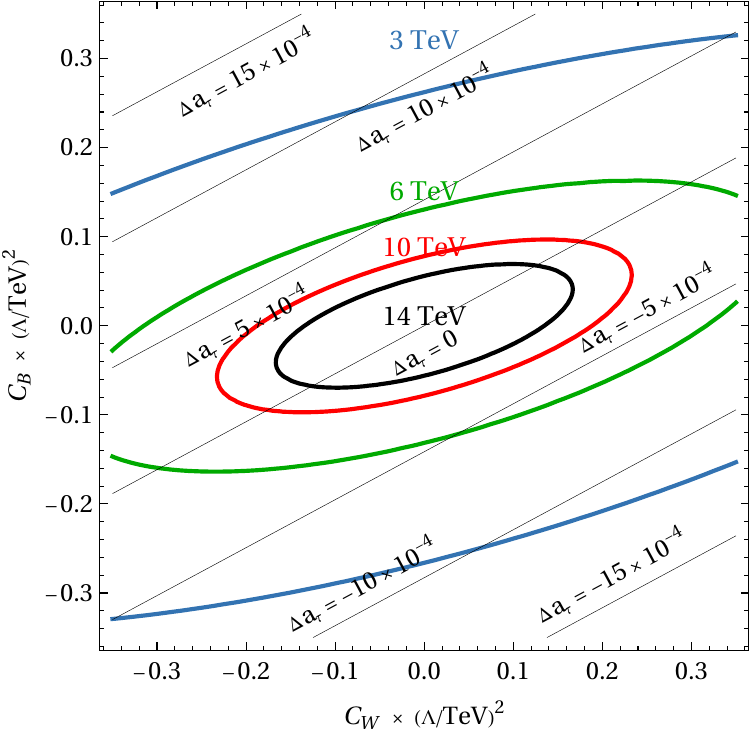}
        \caption*{(c) $\mu^+\mu^-\to \mu^+ \mu^- \tau^+\tau^- $}
    \end{minipage}%
    \hfill%
    \begin{minipage}[b]{0.22\textwidth}
        \centering%
        \includegraphics[width=\textwidth]{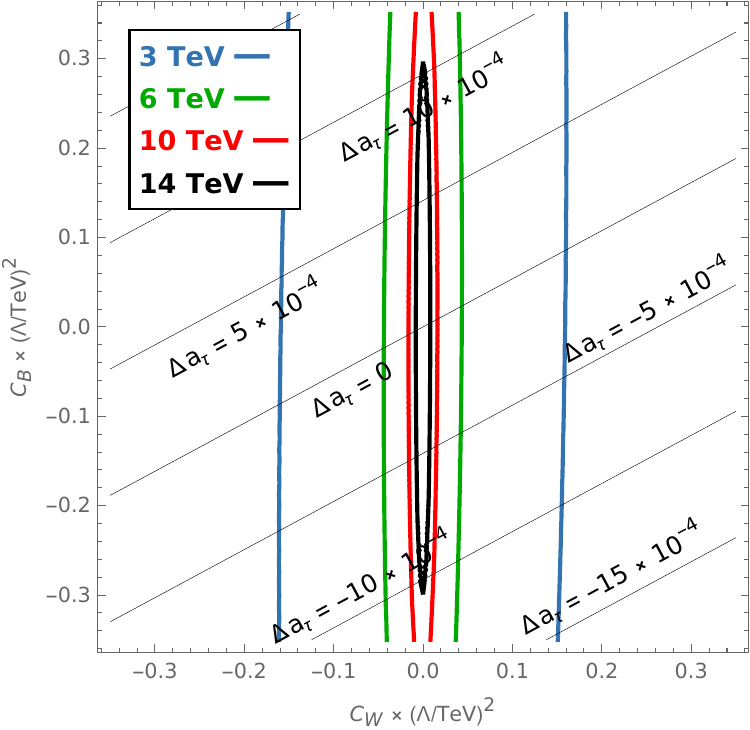}
     \caption*{(d) $\mu^+\mu^-\to \nu_\mu \bar{\nu}_\mu  \tau^+\tau^-$}
    \end{minipage}
       \caption{ 95\% CL limits on the Wilson coefficients $C_{eW}$ and $C_{eB}$ from the Drell-Yan processes 
$\mu^+\mu^- \to \tau^+\tau^-(h)$, or vector-boson-fusion processes such 
as $\mu^+\mu^- \to \mu^+\mu^-\tau^+\tau^-$ and $\mu^+\mu^-\to \nu_\mu \bar{\nu}_\mu  \tau^+\tau^-$ at muon colliders of different energies; the black isolines indicate the corresponding value of $\Delta a_{\tau}$. \label{limits}}
\end{figure}

We compute the significance as $N_{\rm sig}/\sqrt{N_{\rm sig}+N_{\rm bkg}}$.
In the Figure~\ref{limits}, the plots explore how muon colliders with different center-of-mass energies from 3 TeV up to 14 TeV affect the constraints on the Wilson coefficients.  As the energy increases, the contours tighten, indicating that higher-energy colliders provide better sensitivity and stronger constraints on $C_{eW}$ and $C_{eB}$. These constraints could reach sensitivity levels comparable to the tau anomalous magnetic moment $(g\!-\!2)_\tau$ at the level of $\mathcal{O}(10^{-5}\!-\!10^{-4})$, potentially to $\mathcal{O}(10^{-6})$. The results shown here are preliminary and  further details will be presented in our forthcoming paper~\cite{paper:2024}. In summary, future high-energy muon colliders could substantially enhance our ability to probe new physics associated with the anomalous magnetic moment $\Delta a_\tau$.

\end{document}